\documentclass[twocolumn,amsmath,preprintnumbers,amssymb,showpacs]{revtex4}

\usepackage{graphicx}
\usepackage{dcolumn}
\usepackage{bm}
\usepackage[cp1250]{inputenc}
\usepackage{ifthen} 

\begin{document}
\newcommand{\bk}{\mathbf{k}}
\newcommand{\bQ}{\mathbf{Q}}
\newcommand{\mb}{\mathbf}
\newcommand{\eq}{\begin{equation}}
\newcommand{\eqx}{\end{equation}}
\newcommand{\eqn}{\begin{eqnarray}}
\newcommand{\eqnx}{\end{eqnarray}}

\preprint{APS/123-QED}

\title{Superconducting BCS versus Fulde-Ferrell-Larkin-Ovchinnikov states of quasiparticles with spin dependent mass and their distinguishability}

\author{J. Spałek$^1$}
 \email{ufspalek@if.uj.edu.pl}

\author{M. Maśka$^2$}

\author{M. Mierzejewski$^2$}

\author{J. Kaczmarczyk$^1$}

\affiliation{
$^1$Instytut Fizyki im. Mariana Smoluchowskiego, Uniwersytet Jagielloński, Reymonta 4 30-059 Kraków, Poland \linebreak
$^2$Institute of Physics, University of Silesia, Uniwersytecka 4, 40-007 Katowice, Poland
}

\date{\today}

\begin{abstract}
The first observation of the Fulde-Ferrell-Larkin-Ovchinnikov (FFLO) superconducting state and a subsequent detection of the spin-dependent effective masses of quasiparticles in the CeCoIn$_5$ heavy fermion system are combined into a single theoretical framework. The appearance of the spin-split masses extends essentially the regime of temperatures and applied magnetic fields, in which FFLO is observable and thus is claimed to be very important for the FFLO detectability. We also stress that the quasiparticles composing Cooper pair become \textit{distinguishable} in the nonzero field. The analysis is performed within the Kondo-lattice limit of the finite-$U$ Anderson - lattice model containing both the mass renormalization and real-space pairing within a single scheme.
\end{abstract}



\maketitle

The superconductors discovered at the beginning of this decade are termed as \textit{unconventional} because the breakdown of the basic symmetries such as the spatial \cite{1} or time \cite{2} or the spin singlet-triplet Cooper pair mixing \cite{3} have been discovered in the heavy-fermion compounds \cite{4}. These features alone lead to a \textit{nonstandard} behavior even when the Bogolyubov-de Gennes quasiparticle approach is applicable. Such an approach is based on the concept of Landau-Fermi liquid, albeit almost localized \cite{5}. A separate, largely unanswered question, is concerned with the pairing in the non-Fermi (non-Landau) liquids \cite{6} and particularly, in high temperature superconductors \cite{7}. In this situation, it seems proper to select cases, for which a limited number of novel factors such as the spin-dependence of quasiparticle mass \cite{8} and the real-space pairing induced by strong correlations \cite{9}, can be incorporated into an effective Fermi-liquid picture, and thus become tractable theoretically \cite{10}. We suggest, that the heavy-fermion systems such as CeCoIn$_5$ \cite{11} and the specific organic compounds \cite{12} can be selected as such test cases.

In this paper we take into account spin dependent quasiparticle masses \cite{8} observed recently \cite{13} and analyze in detail the relative stability of the Bardeen-Cooper-Schrieffer (BCS) phase \cite{14} against the Fulde-Ferrell-Larkin-Ovchinnikov (FFLO) state \cite{15} in an applied magnetic field $\mathbf{H}$ and at temperature $T \geq 0$. The basic question we tackle, apart from a detailed quantitative analysis, is why the FFLO phase became so easily observable in this strongly correlated system, after almost forty years of experimental search for that phase existence. By introducing the spin-split masses $m_\sigma$ ($\sigma = \pm1$ is the quasiparticle spin quantum number), one deals, in our view, with a fundamentally new situation, as by switching on the applied field one may transform the system of quantum mechanically \textit{indistinguishable quasiparticles} into their \textit{distinguishable} correspondants. This circumstance produces not only changes of the normal-state properties such as the appearance of itinerant-electron metamagnetism \cite{16}, but also leads to the essential modification of the single Cooper-pair bound state composed of two \textit{distinguishable} particles \cite{17}. In this manner, the spin-dependent renormalization of the quasiparticle mass introduces by itself a non-Landau feature to fermionic liquid and appears only when the correlated electrons are treated within a non-perturbational approach \cite{8, 16, 18}. Here we concentrate first on the condensed state of the pairs, which retain their \textit{indistinguishability} and discuss subsequently the properties of a single Cooper pair, with partners becoming \textit{distinguishable} in the quantum mechanical sense, in the applied field.

We start with the effective Anderson-lattice model in the large but finite-$U$ limit \cite{18}, which has the form

\begin{eqnarray}
H & = & \sum_{m n} ( t_{m n} - \mu \delta _{m n} ) c_{m \sigma}^\dagger c_{n\sigma} + \epsilon_f \sum_{i \sigma} N_{i \sigma} (1-N_{i \overline{ \sigma } }) \nonumber  \\
& + & \sum_{i \sigma } V_{im} N_{i \sigma } (1-N_{i \overline{ \sigma } }) (f_{i\sigma }^\dagger c_{m \sigma } + c_{m\sigma }^\dagger f_{i\sigma }) \nonumber \\
 & - & 2 \sum_{i m n} \frac{2 V_{i m} V_{i n}}{ U + \epsilon_f } b_{im}^\dagger b_{in}. \label{eq:H1}
\end{eqnarray}

The first term represents the conduction ($c$) electrons, the second the originally localized ($f$) electrons with the site double occupancies projected out ($N_{i\sigma} \equiv f_{i\sigma}^\dagger f_{i\sigma}$), the third - the projected $f$ - $c$ hybridization, and the fourth the real-space pairing part in the leading order, with the projected spin-singlet pairing operators

\begin{equation}
b^\dagger_{im} = \frac{1}{\sqrt{2}} [a^\dagger_{i\uparrow} (1-N_{i\downarrow}) c_{m\downarrow} - a^\dagger_{i\downarrow}(1-N_{i\uparrow})c_{m\uparrow}  ].
\end{equation}

Note that the pairing part disappears if the $f$-$f$ intraatomic interaction $U \rightarrow \infty$. The crucial features of (\ref{eq:H1}) are the inclusion of the projection which leads to the spin-dependent renormalization of the hybridization and in turn, the spin-split quasiparticle masses, as well as of the virtual ($f$-$c$) hopping induced real-space pairing, both within a single scheme. In the \textit{Kondo-lattice limit} i.e. when $n_f \equiv \langle \sum_\sigma N_{i\sigma} (1-N_{i\sigma}) \rangle \equiv 1-\delta$, with $\delta \ll 1$) and in the saddle-point or Gutzwiller approximations \cite{5}, the Hamiltonian (\ref{eq:H1}) can be brought (up to a constant term) into a single-narrow band form with the BCS-type of pairing \cite{9} including the Zeeman term (field $H\neq 0$)

\begin{eqnarray}
\nonumber & \mathcal{H} & = \sum_{\bk \sigma} (q_\sigma \epsilon_\bk - \sigma \mu_B H) \psi^\dagger_{\bk \sigma} \psi_{\bk \sigma} - \\
 &&\frac{1}{N} \sum_{\bk \bk'}
 \frac{4V_\bk^2 V_{\bk'}^2}{\epsilon_f^2 (\epsilon_f + U)}R_{\sigma\overline{\sigma}}
 \gamma_\bk \gamma_{\bk'} \psi^\dagger_{\bk \uparrow} \psi^\dagger_{-\bk \downarrow} \psi_{-\bk' \downarrow} \psi_{\bk' \uparrow},
\end{eqnarray}

where the operators $\psi^\dagger_{\bk \sigma} \approx f^\dagger_{\bk \uparrow}$, $q_\sigma \epsilon_\bk$ being the band energy of itinerant $f$ electrons induced by the hybridization, $\gamma_\bk$ is a complicated function \cite{9} of quasimomentum $\bk$, and $V_{\bk\bk'} = -4 V_\bk^2 V_{\bk'}^2/(\epsilon_f^2 (\epsilon_f + U))$ is the strength of the pairing potential, taken in the following for intraatomic form of hybridization, $V_\bk = V$. The factor $q_\sigma \simeq (1-n_f)/(1-n_{f\sigma})$ is the hybridization renormalization implying the spin dependent mass renormalization, $m_\sigma/m_0 = q_\sigma^{-1}$. Physically, the pairing results from the Kondo-type self-screening of the heavy quasiparticles. The renormalization factor of the pairing part $R_{\sigma\overline{\sigma}} \sim (q_{\sigma}q_{\overline{\sigma}})^{1/2}$ will be regarded as a constant reducing the pairing potential magnitude $V_0 \equiv \frac{4 V^4}{\epsilon_f^2 (\epsilon_f + U)}$. In what follows we take the form of $\gamma_\bk$ as the $d-$wave in the two-dimensional limit.

A few important physical remarks conveying principal features of our approach are in place here. First, the original atomic $f$ electrons acquire band properties by a three-step process: hopping $f\rightarrow c$ from the atomic to the bare conduction band states, followed by a propagation in the conduction band, and a subsequent deexcitation $c\rightarrow f$. Hence, the hopping amplitude for $f$ states is of the order $(V/\epsilon_f)^2t_{mn}$. Second, the mechanism of pairing is of analogous origin as that in t-J model with a different renormalization factors for the pairing part \cite{19}. Third, the strongly correlated and hybridized electrons form an almost localized Fermi liquid \cite{10} of electrons in a very narrow band of $f$ states. The last assumption means that the number of quasiparticles characterized by $n_f$ (related to e. g. cerium valency by Ce$^{+4-n_f}$) is constant as a function of $T$ and $H$ which is valid only at low temperature $T \ll T_K$ ($T_K$ is the effective Kondo temperature \cite{9}). Finally, $T_K$ plays the role of the $f$-electron bandwidth and its value is of the order $2 z |t| (V/\epsilon_f)^2(1-n_f) \sim 10^{-2} t \equiv z \tilde{t}$, where $t$ is the amplitude of the hopping $t_{mn}$ between $z$ nearest neighbors. Since $t$ is a fraction of $eV$, then $T_K$ is in the regime $10-100\, K$. This is the reason why the heavy electrons are so sensitive to Zeeman fields of the order of $10\,T$.

We discuss now the relative stability of the BCS and the FFLO condensed states, the latter defined by the center-of-mass momentum $\bQ \neq 0$ of the Cooper pairs and by the gap amplitude $\Delta_\bQ$. An essential factor is that the states for $\mathbf{H} \neq 0$ have spin-dependent masses $m_\sigma$, as the field acts on both quasiparticles resulting from the broken pairs, as well as from gapless regions of $\Delta_\mathbf{k} \equiv \Delta_{\mathbf{k}, \bQ}$. Having in mind different gap symmetries, we assume that $\Delta_\bk \equiv \Delta_\bQ \gamma(\bk)$, where the symmetry expressed via $\gamma(\bk)$ is superposed on $\bQ$ dependence of the gap magnitude. We analyze explicitly the case of quasi-two-dimensional superconductor \cite{20} in a clean limit, where Pauli-limiting situation \cite{21} is well defined. Also, as we consider real space pairing caused by strong correlations, we require a detailed dispersion relation in the full band, which is parametrized as follows

\eq
\epsilon_{\bk\sigma} = q_\sigma [-2 t (\cos k_x + \cos k_y) + 4 t' \cos k_x \cos k_y ],
\eqx

with $t$ and $t'$ being hopping integrals between the nearest and the next-nearest neighbors (we take $t'/t = 0.5$). This electronic structure represents properly the situation in both CeCoIn$_5$ \cite{22} and of $\kappa-(BEDT-TTF)_2 Cu(NCS)_2$. Using the standard diagonalization techniques we obtain the quasiparticle energies in the superconducting state in the form similar to that in \cite{20, 23}, i.e.

\eqn
\nonumber E_{\bk, \bQ, \pm} & = & g \mu_B H + \frac{1}{2} (\epsilon_{\bk \uparrow} - \epsilon_{-\bk+\bQ \downarrow}) \\
& \pm & \frac{1}{2} [(\epsilon_{\bk \uparrow} + \epsilon_{-\bk+\bQ \downarrow}-2\mu)^2 + 4|\Delta_{\bk, \bQ}|^2]^{\frac{1}{2}},
\eqnx

where $\mu$ is the chemical potential and the gap is determined from the self-consistent equation

\eq
\Delta_{\bQ} = \frac{V_0 \Delta_{\bQ}}{N} \sum_\bk \gamma_\bk^2\frac{f(E_{\bk, \bQ+}) - f(E_{\bk, \bQ-}) }{E_{\bk, \bQ+} - E_{\bk, \bQ-}}.
\eqx

\noindent This equation must be supplemented by the corresponding equations for $\mu$ and $n_\sigma$. In effect, one has to minimize the free-energy functional $\mathcal{F} = \mathcal{F}(T, H; \Delta_\bQ, \bQ, n_\sigma)$ of the form

\eqn
\nonumber \mathcal{F} & = & - k_B T \sum_\bk \sum_{\alpha = \pm} ln[1+\exp(-E_{\bk, \bQ, \alpha}/k_B T)] + \mu N n_f\\
& + & \sum_\bk (\epsilon_{-\bk-\bQ \downarrow} - \mu - g \mu_B H) + N |\Delta_\bQ|^2/V_0,
\eqnx

 (N being the number of sites), with respect to $\bQ$, $\Delta_\bQ$, and the band fillings $n_\sigma$ (subject to the constraint $n_\sigma + n_{ \overline{\sigma} } = n_f$). The solutions were determined by the two methods to check for the accuracy: first by sampling the Brillouin zone and second by integrating over bare density of states $\rho_0(\epsilon)$ with a variable grid depending on the steepness of $\rho_0(\epsilon)$ behavior, particularly near the van Hove singularity.

\begin{figure}
\scalebox{0.2}{\includegraphics{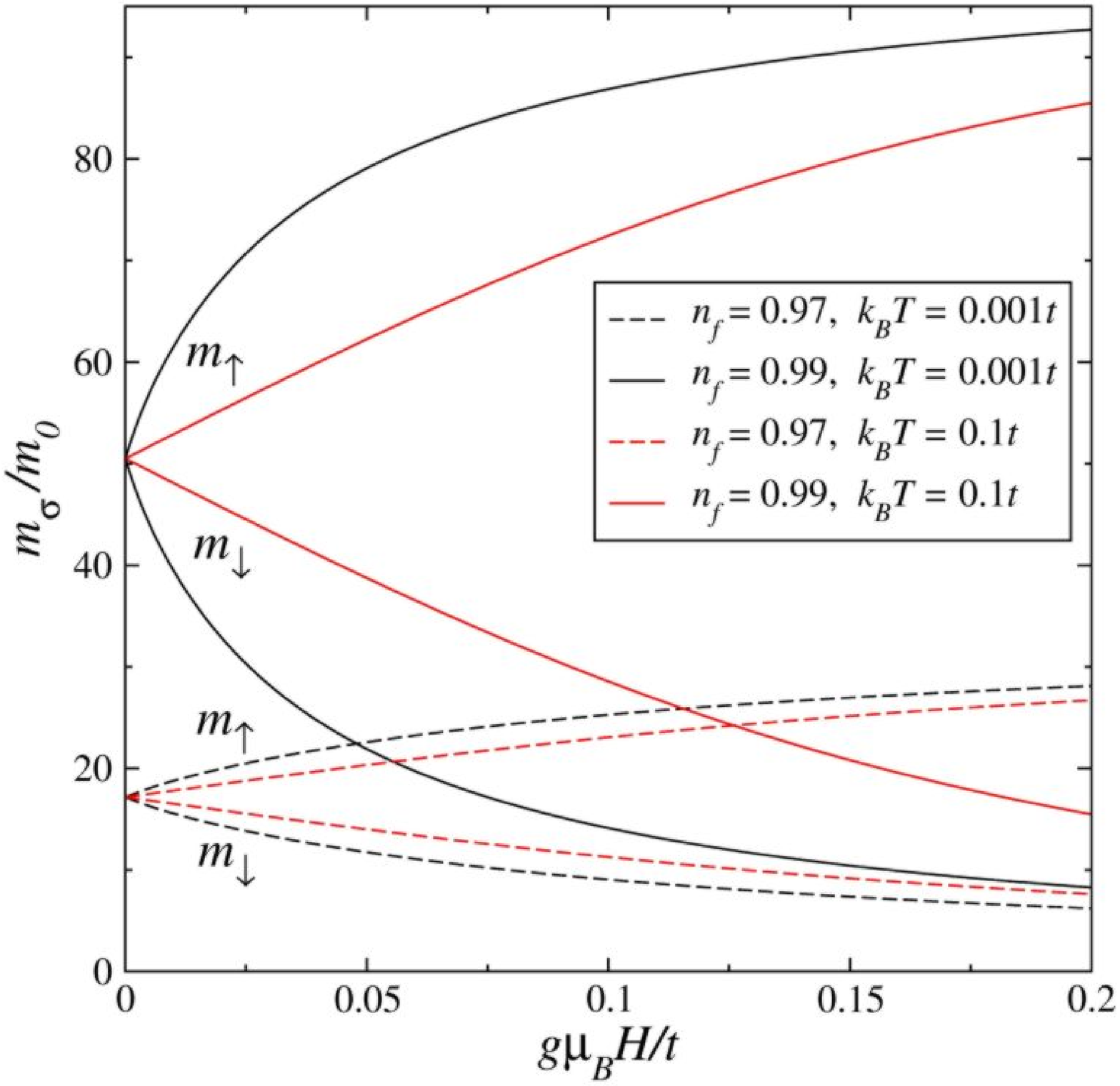}}
\caption{\label{fig:fig1} (Color online). Spin-dependent mass enhancement vs. $H$ for selected values of both the filling $n_f$ and temperature $T$. The effect is particularly strong for $n_f$ very close to unity and for $T \ll T_K$.}
\end{figure}

In Fig. 1ab we display the field dependence of the mass enhancement in the normal state close to the integer $n_f$. Note a strong temperature dependence associated with the reduction of the magnetization ($n_\uparrow - n_\downarrow$). Taking $\tilde{t} = 20\,K$, we estimate the regime of physical fields $H<20 \, T$ as limited by $\mu_B H/t \leq 0.1$, where the mass splitting is indeed essential.

\begin{figure}
\scalebox{0.2}{\includegraphics{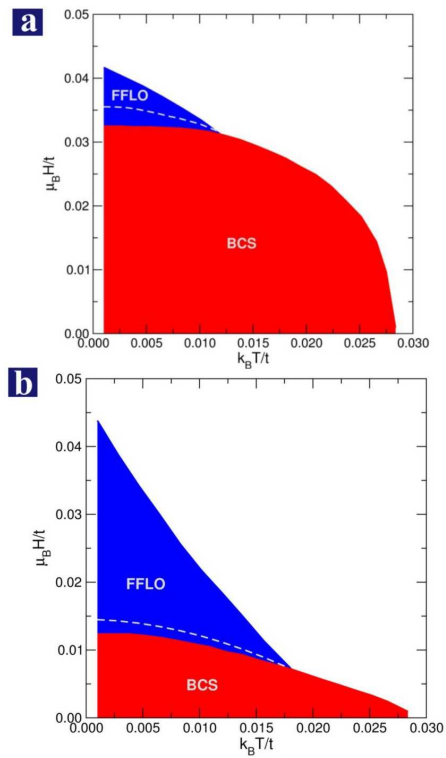}}
\caption{\label{fig:fig2} (Color online). Phase boundaries for a $d$-wave superconductor with both the spin-independent masses (a) and with the spin-split masses (b). The FFLO-BCS transition line is discontinuous. The dashed line marks the stability limit of the BCS state as determined by the value of the critical field $H_{c2}$. The values of parameters: $n_f = 0.97$ and $V_0 = 10 \, K$.}
\end{figure}

To characterize the stability of the BCS and FFLO states, we plot the phase diagrams in Fig. 2ab with spin-independent (average $m_{av} = (m_\uparrow + m_\downarrow)/2$) and with the spin-dependent ($m_\sigma \neq m_{ \overline{ \sigma } }$) masses, respectively. One notes the essential extension of the stability regime (of the FFLO phase) on the $H-T$ plane in the latter case. This is the reason, we claim, why the FFLO state has been observed for the first time in CeCoIn$_5$, where the spin-dependent masses were observed \cite{13}. For completeness, in Fig. 3ab we plot the values of the gap $\Delta_\bQ$ and the optimal $\bQ$ values, respectively ($\Delta_{\bQ = 0}$ in the BCS state and $\Delta_0 \equiv \Delta$ then). The order parameter $\Delta$ jumps across the BCS-FFLO border signalling the 1st-order transition. The transition to the normal phase seems to be quasicontinuous. The dashed lines in Fig. 2ab mark the critical field $H_{c2}$ for the BCS state. Also, the uppermost critical field is spectacularly curved upward only if we include the spin mass splitting, which is observed in the organic system \cite{23}, as well as the discontinuous nature of the BCS-FFLO transition.

\begin{figure}
\scalebox{0.2}{\includegraphics{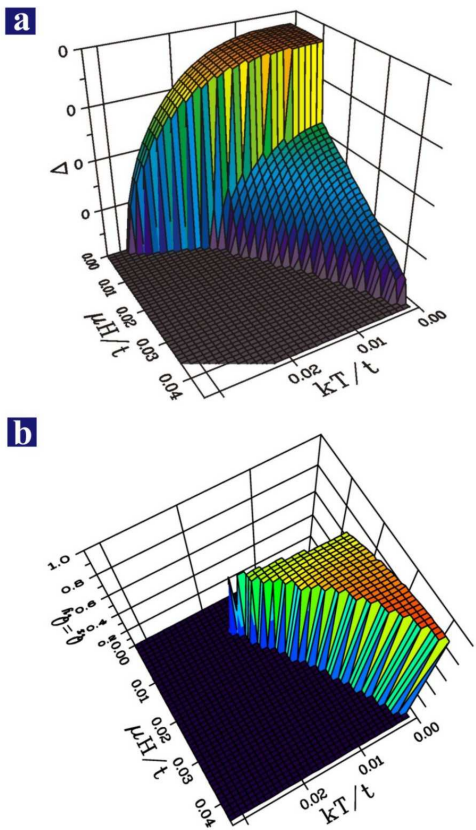}}
\caption{\label{fig:fig3} (Color online). Magnitude $\Delta$ of the order parameter on the $H-T$ plane (a) and the center-of-mass momentum $\bQ$ (b) for the same situation as above in the case with spin-dependent masses.}
\end{figure}

One should note that the free-energy differences at the transition are of the order of $10^{-4} t$ at most, so the accuracy of extracting the solution of the self-consistent integral equations for $\Delta_\bQ$, $n_\sigma$, and $\mu$ is particularly intricate. This circumstance is also the reason behind the statement that within our numerical accuracy, we cannot decisively state that the FFLO-normal state phase transition is continuous. The last remark is illustrated in Fig. 3ab, where we plot the gap amplitude on the $H-T$ plane. In Fig. 3b we provide the value of $\bQ$, which is a well defined quantity in FFLO regime ($Q_x = Q_y \approx 0.2-0.8$ in $\pi/a$ units, where $a$ is the lattice constant).

From the results above it follows that the spin-dependent masses play an important quantitative role in the stability of the phases. They lead also to a qualitative, \textit{if not fundamental}, change of the description of individual particles (cf. Fig. 1) and of the single-Cooper pair state. Namely, since the mass splitting $m_\uparrow - m_\downarrow \sim (n_{f\uparrow} - n_{f\downarrow})$, the \textit{indistinguishable} particles in the field (or spin-polarization) absence, become \textit{distinguishable} for $H>0$. To demonstrate this we note first, that since the masses are labeled by the spin quantum number, this means that the spin part of the wave function should be chosen either as $\chi_\uparrow(1)\chi_\downarrow(2)$ or as $\chi_\downarrow(1)\chi_\uparrow(2)$, not in the usual singlet form $[\chi_\uparrow(1)\chi_\downarrow(2)-\chi_\downarrow(1)\chi_\uparrow(2)]/\sqrt{2}$. This is what we mean \cite{25} by the \textit{quasiparticle distinguishability}. To test our concepts we reformulated \cite{16, 26} the original Cooper problem including the momentum $\bQ \neq 0$. The basic Cooper-pair properties are displayed in Fig. 4ab. In Fig. 4a we plot the pair binding energy $\Delta$ in a few situations: the uppermost solid curve is the dependence $\Delta_\bQ(H)$ with the spin dependent masses, starting from $m_{av} = 304 \, m_0$ and $\bQ \neq 0$; the solid line just below reflects the corresponding dependence for $\bQ = 0$. The modified spin-wave function essentially increases the binding energy \cite{16}. In Fig. 4b we plot the value of $|\bQ|$ vs. $H$. The jump to the value $Q \approx |k_{F\uparrow} - k_{F\downarrow}|$ ($k_{F\sigma}$ denotes the Fermi wavevector for the $\sigma$-subband) is represented by the empty circle in Fig. 4a. The solid circles define the field values destabilizing the pair bound state through the spin flip. The stable state of the single-pair state with $Q \neq 0$ in $H > 2.5 \, T$ is thus the prerequisite of the FFLO state.

\begin{figure}
\scalebox{0.18}{\includegraphics{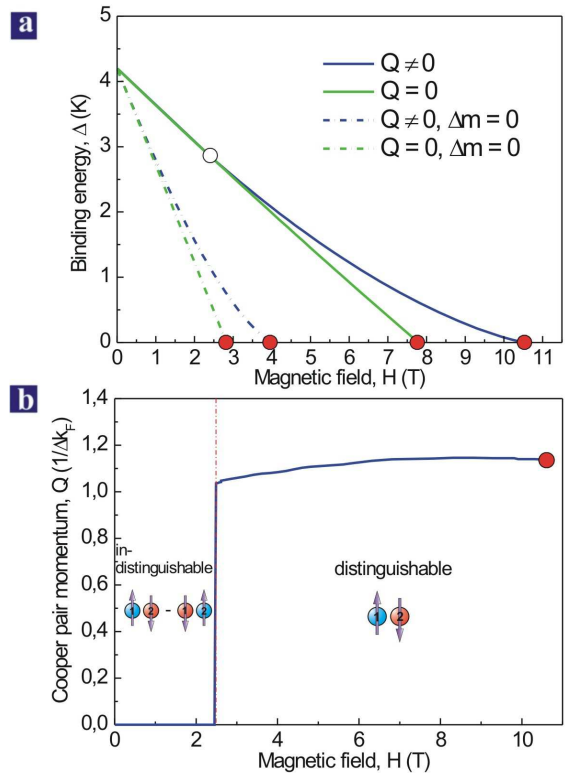}}
\caption{\label{fig:fig4} (Color online). (a) field dependence of single-pair binding energy with spin-split masses ($\Delta m \neq 0$) and without ($\Delta m = 0$), for both center of mass momentum $\bQ \neq 0$ and $\bQ = 0$. The empty circle marks the transition to the state with $\bQ$ roughly equal to the difference of the Fermi momenta $k_{F\uparrow}$ and $k_{F\downarrow}$, whereas those on the abcissa are the Pauli fields for the bound-state destruction. (b) $\bQ$ vs. H for the stable pair state (the spin wave-function symmetry is marked).}
\end{figure}


In summary, we have applied the concept of the spin-direction dependent mass for a fermionic liquid composed of quasiparticles that condense into the BCS or FFLO states. This basic property extends the regime of the FFLO phase stability on the expense of the BCS state. The FFLO regime is stabilized by the circumstance that the mass difference enhances behavior of system corresponding difference in the density of states $\rho_\sigma(\epsilon)$. We also point out to the basic nontrivial feature of the quasiparticles composing Cooper pair. Namely, we can study an evolution of the two-particle state from the limit of \textit{indistinguishable to distinguishable} quantum particles as a function of physically controllable parameter.
Such result appears in both the Gutzwiller (or mean field) and the dynamic mean-field levels \cite{8, 17}. The situation close to the metamagnetic point requires a separate analysis of an effective field appearing on the saddle-point level (cf. \cite{8}).

The authors acknowledge Grants from the Ministry of Science and Higher Education. This work was performed under the Grant COST-P16 "\textit{Emergent Behaviour in Correlated Matter}" from the European Science Foundation. Useful discussions with Ilya Sheikin at SCES'2008 and with Frank Steglich, are appreciated.

{}

\end{document}